\documentclass[preprintnumbers,prl,twocolumn,showpacs,floatfix,amsmath,amssymb,superscriptaddress]{revtex4}
\usepackage{bm}
\usepackage{graphicx}
 
\newcommand{\ud}{\mathrm{d}} 
 
\newcommand{\GF}{G_\mathrm{F}}
 
\newcommand{\Hb}{\mathbf{H}}

\newcommand{\sB}{\mathbf{s}}

\newcommand{\Lm}{\mathcal{L}}
\newcommand{\ntot}{n_\nu^\mathrm{tot}}

\newcommand{\HV}{\Hb_\mathrm{V}}
\newcommand{\EC}{E_\mathrm{C}}
\newcommand{\wc}{\omega_\mathrm{pr}}

\newcommand{\Esync}{E_\mathrm{sync}}
\newcommand{\basef}[1]{\bm{\hat{\mathbf{e}}}^\mathrm{f}_{#1}}

\newcommand{\thetav}{\theta_\mathrm{v}}

\newcommand{\myfigsep}{0.03 \textwidth}
\newcommand{\myfigwid}{0.41 \textwidth}

\begin{document}
\title{Neutrino Mass Hierarchy and Stepwise Spectral Swapping of 
Supernova Neutrino Flavors}
\newcommand*{\UCSD}{Department of Physics, %
University of California, San Diego, %
La Jolla, CA 92093-0319}
\affiliation{\UCSD}
\newcommand*{\LANL}{Theoretical Division, Los Alamos National Laboratory, %
Los Alamos, NM 87545}
\affiliation{\LANL}
\newcommand*{\UMN}{School of Physics and Astronomy, %
University of Minnesota, Minneapolis, MN 55455}
\affiliation{\UMN}

\author{Huaiyu Duan}
%\email{hduan@ucsd.edu}
\affiliation{\UCSD}
\author{George M.~Fuller}
%\email{gfuller@ucsd.edu}
\affiliation{\UCSD}
\author{J.~Carlson}
%\email{carlson@lanl.gov}
\affiliation{\LANL}
\author{Yong-Zhong Qian}
%\email{qian@physics.umn.edu}
\affiliation{\UMN}

\date{\today}

\begin{abstract}
We examine a phenomenon recently predicted by numerical  simulations
of supernova neutrino 
flavor evolution: the swapping of supernova $\nu_e$ and 
$\nu_{\mu,\tau}$ energy spectra below (above) energy $\EC$ for the 
normal (inverted) neutrino mass hierarchy. We present the results of 
large-scale numerical calculations which show that in the normal neutrino 
mass hierarchy case, $\EC$ decreases as the assumed 
effective $2\times 2$ vacuum $\nu_e\rightleftharpoons\nu_{\mu,\tau}$ 
mixing angle ($\simeq \theta_{1 3}$) is decreased. In contrast, these 
calculations indicate that $\EC$ is essentially independent of 
the vacuum mixing angle in the inverted neutrino mass hierarchy case. 
With a good neutrino signal from a future Galactic supernova, the above
results could be used to determine the neutrino mass hierarchy
even if $\theta_{13}$ is too small to be measured
by terrestrial neutrino oscillation experiments.

\end{abstract}


\maketitle

In this letter we point out how two grand themes in contemporary science, 
physics beyond the Standard Model of elementary particles and the physics 
of stars undergoing gravitational collapse, overlap in a way that could 
allow unique insight into the nature of neutrinos. Recent experiments have 
established that neutrinos have non-vanishing rest masses and that the 
flavor states $\nu_e$, $\nu_\mu$ and $\nu_\tau$ for these particles are 
mixtures of the vacuum mass eigenstates $\nu_1$, $\nu_2$ and $\nu_3$
(see, \textit{e.g.}, Refs.~\cite{PDBook,Strumia:2006db} for
recent reviews). 
However, key issues remain unresolved. Among these is the nature of 
the neutrino mass hierarchy: the sign of the mass-squared difference 
$\delta m^2_{31} = m_3^2-m_1^2\simeq \pm \delta m^2_\mathrm{atm}$ 
remains unknown. Here $\delta m^2_\mathrm{atm}$ is the neutrino 
mass-squared difference associated with atmospheric neutrino oscillations, 
and the plus (minus) sign corresponds to the 
normal (inverted) neutrino mass hierarchy.
Conventional laboratory experimental resolution of the mass hierarchy 
issue is problematic, in part because neutrino rest masses are tiny 
and because $\theta_{13}$, the mixing angle relating $\nu_e$ to $\nu_3$,
is small. 
One possible way to probe the neutrino mass
hierarchy is to analyze neutrino signals from Galactic 
supernovae.  Supernova neutrinos can experience significant
flavor transformation through the Mikheyev-Smirnov-Wolfenstein (MSW) 
effect \cite{Wolfenstein:1977ue,Mikheyev:1985aa} as they
stream out from the surface of a proto-neutron star (with very high
matter density) into the vacuum. In addition, it has been pointed
out that neutrino-neutrino forward scattering can provide
an additional source for neutrino refractive indices
\cite{Fuller:1987aa,Notzold:1988kx,Pantaleone:1992xh,Sigl:1992fn}.
This neutrino self-coupling is especially important in
the supernova environment because neutrino fluxes are large.
Therefore, previous studies of supernova neutrino oscillations 
based on the pure MSW effect
\cite{Dighe:1999bi,Schirato:2002tg,%
Lunardini:2003eh,Kneller:2005hf,Chiu:2006zs,Yoshida:2006qz,Kneller:2007kg}
may not apply in some scenarios.
Following suggestions in Refs.~\cite{Duan:2006an,Duan:2007mv,Raffelt:2007cb},
in this letter we present and analyze new large-scale numerical
calculations of supernova neutrino flavor evolution that suggest
a novel method to determine the neutrino mass hierarchy.
This method is independent of absolute neutrino masses and
can work even for tiny $\theta_{13}$.

Our method is based on a stunning
feature revealed by recent  numerical simulations of
supernova neutrino flavor evolution \cite{Duan:2006jv,Duan:2006an}:
(1) For the normal neutrino mass hierarchy case,
$\nu_e$ and $\nu_{\mu,\tau}$ swap their energy spectra
at energies below a transition energy $\EC$, but 
retain their original spectra at higher energies;
(2) For the inverted neutrino mass hierarchy case, the situation is
exactly the opposite. This phenomenon is known as 
``stepwise spectral swapping'' \cite{Duan:2006an}
or ``spectral split'' \cite{Raffelt:2007cb}.

The stepwise swapping of the $\nu_e$ and $\nu_{\mu,\tau}$
energy spectra has its origin in nonlinear neutrino self-coupling.
Assuming coherent neutrino propagation and the efficacy 
of the mean field approach  \cite{Friedland:2003eh,Balantekin:2006tg},
for $2\times2$ flavor evolution, it is possible to 
define the neutrino flavor isospin (NFIS) as 
 \cite{Duan:2005cp}
%$\sB_\nu\equiv\psi_\nu^\dagger(\bm{\sigma}/2)\psi_\nu$
%and $\sB_{\bar\nu}\equiv(\sigma_y\psi_{\bar\nu})^\dagger(\bm{\sigma}/2)(\sigma_y\psi_{\bar\nu})$
\begin{equation}
\sB_\nu\equiv\psi_\nu^\dagger\frac{\bm{\sigma}}{2}\psi_\nu
\quad\text{and}\quad
\sB_{\bar\nu}\equiv(\sigma_y\psi_{\bar\nu})^\dagger\frac{\bm{\sigma}}{2}
(\sigma_y\psi_{\bar\nu})
\end{equation}
for a neutrino (with flavor
wavefunction $\psi_\nu$) and an antineutrino ($\psi_{\bar\nu}$),
respectively.
Flavor evolution for neutrino or antineutrino mode $i$ is
described by precession of a corresponding NFIS $\sB_i$ around an
effective field:
\begin{equation}
\begin{split}
\frac{\ud}{\ud t}\sB_i 
 &
= \sB_i\times
\Big[\omega_i\HV-\sqrt{2}\GF n_e\basef{z}
 \\ &\quad
-2\sqrt{2}\GF\sum_j(1-\cos\vartheta_{ij})n_j\sB_j\Big].
\end{split}
\label{eq:eom}
\end{equation}
Here $\basef{x,y,z}$ are the flavor-basis unit-vectors
in flavor space, $\HV\equiv-\sin2\thetav\basef{x}+\cos2\thetav\basef{z}$
generates vacuum mixing for a nonvanishing effective mixing angle $\thetav$, 
$\omega_i=\pm\delta m^2/2E_i$ is the vacuum
precession angular velocity
around $\HV$ for a NFIS corresponding to
 a neutrino (plus sign) or antineutrino (minus sign)
with energy $E_i$, 
$\GF$ is the Fermi constant,
$n_e$ is the net electron number density, $\vartheta_{ij}$ is the angle
between the propagation directions of neutrinos in modes $i$ and $j$,
and $n_j$ is the number density of neutrinos in mode $j$. Because
flavor transformation in the $\nu_e\rightleftharpoons\nu_{\mu,\tau}$
and $\bar\nu_e\rightleftharpoons\bar\nu_{\mu,\tau}$ channels
is the most important in supernovae (\textit{e.g.}, 
for shock reheating \cite{Fuller:1992aa,Fuller:2005ae} and nucleosynthesis 
\cite{Qian:1993dg,Pastor:2002we,Balantekin:2004ug,Fuller:2005ae}), 
and because $\delta m^2_\mathrm{atm}$ will give flavor
transformation deeper in the supernova envelope  than
will $\delta m^2_\odot$, the mass-squared difference
associated with solar neutrino oscillations, we take 
$\delta m^2=\pm3\times10^{-3}\,\mathrm{eV}^2
\simeq\pm\delta m^2_\mathrm{atm}$ and
$\thetav\simeq\theta_{13}\ll1$.
For this $2\times2$ mixing we use $\nu_{\tau^*}$ to
designate the relevant linear combination of $\nu_\mu$ and 
$\nu_\tau$ \cite{Balantekin:1999dx}.

In a stepwise-swapping scenario, 
the probability $P_{\nu\nu}$ for neutrinos to remain 
in their initial flavor states
is
\begin{equation}
P_{\nu\nu}(\omega)\simeq \frac{1}{2}
\left[1-\mathrm{sgn}(\omega-\wc^0)\right],
\label{eq:step}
\end{equation}
where 
$\mathrm{sgn}(\xi)=\xi/|\xi|$ is the sign of $\xi$, and
$\wc^0=\delta m^2/2\EC$ specifies the transition energy $\EC$. 
Eq.~\eqref{eq:step} is slightly different from
Eq.~(57b) in Ref.~\cite{Duan:2007fw}, with different conventions
for $\thetav$ and $\delta m^2$.
This stepwise spectral swapping feature has been
demonstrated in two different approaches
\cite{Duan:2006jv,Duan:2006an,Esteban-Pretel:2007ec,Fogli:2007bk}:
``multi-angle'' simulations, where flavor evolution on 
independently-followed neutrino 
trajectories is self-consistently coupled,
and  ``single-angle'' 
simulations, where the evolution history of radially propagating neutrinos is 
assumed to apply to all trajectories. 
Single-angle calculations capture
 the qualitative features of
stepwise spectral swapping, and they suggest
the following generic explanation for this phenomenon.

Because neutrinos and antineutrinos are in flavor eigenstates
when they leave the neutrino sphere, they naturally form
a ``bipolar system'' in which the corresponding NFIS's form
two oppositely oriented groups \cite{Duan:2005cp}. 
(Note that $\nu_e$/$\bar\nu_{\tau^*}$
and $\bar\nu_e$/$\nu_{\tau^*}$ correspond to NFIS's in the directions
of $+\basef{z}$ and $-\basef{z}$, respectively.)
This bipolar system behaves like a gyroscopic pendulum in 
flavor space \cite{Hannestad:2006nj}, which can have both 
nutation and precession modes. 
These neutrinos and 
antineutrinos initially follow a quasi-static,  MSW-like solution 
near the neutrino sphere before being driven away from this
solution by the collective nutation of the gyroscopic pendulum
\cite{Duan:2007fw}. Subsequently, the gyroscopic pendulum 
can execute regular precession 
around $\HV$, corresponding to the collective precession of
NFIS's in flavor space \cite{Duan:2007mv}. This
precession, although not perfectly regular, is indeed found in
both single-angle and multi-angle simulations
\cite{Duan:2006an,Duan:2007mv}.
If NFIS's stay in the regular collective precession mode,
a stepwise spectral swapping given by Eq.~\eqref{eq:step}
will occur when the neutrino fluxes decrease toward 0
\cite{Duan:2006an,Duan:2007mv,Raffelt:2007cb}. 
In this case, $\wc^0$ in Eq.~\eqref{eq:step} is just
the precession angular velocity for vanishing 
neutrino fluxes.
 
Strictly speaking, the regular precession mode obtains for
$n_e=0$ where the ``lepton number''
\begin{equation}
\Lm\equiv\int_0^\infty
[f_{\nu_1}(E)-f_{\nu_3}(E)-f_{\bar\nu_1}(E)+f_{\bar\nu_3}(E)]\,\ud E
\label{eq:Lm}
\end{equation}
is conserved \cite{Hannestad:2006nj}. Here $f_{\nu_1(\bar\nu_1)}(E)$ and
$f_{\nu_3(\bar\nu_3)}(E)$ are the distribution functions specifying the
populations of the corresponding neutrino (antineutrino) vacuum mass 
eigenstates within energy interval $\ud E$. These are normalized
to the total (summing over all states) neutrino number density $\ntot$ and
in general evolve with time. The conservation of $\Lm$ 
is exact for $n_e=0$, and holds even when neutrino number densities
change \cite{Duan:2007mv}. Because the presence of the matter field does
not change the collective precession qualitatively 
\cite{Duan:2007mv,Duan:2007fw}, the conservation of $\Lm$ 
can be used to compute $\EC$ 
\cite{Raffelt:2007cb}.

\begin{figure*}
\begin{center}
$\begin{array}{@{}c@{\hspace{\myfigsep}}c@{}}
\includegraphics*[scale=0.17, keepaspectratio]{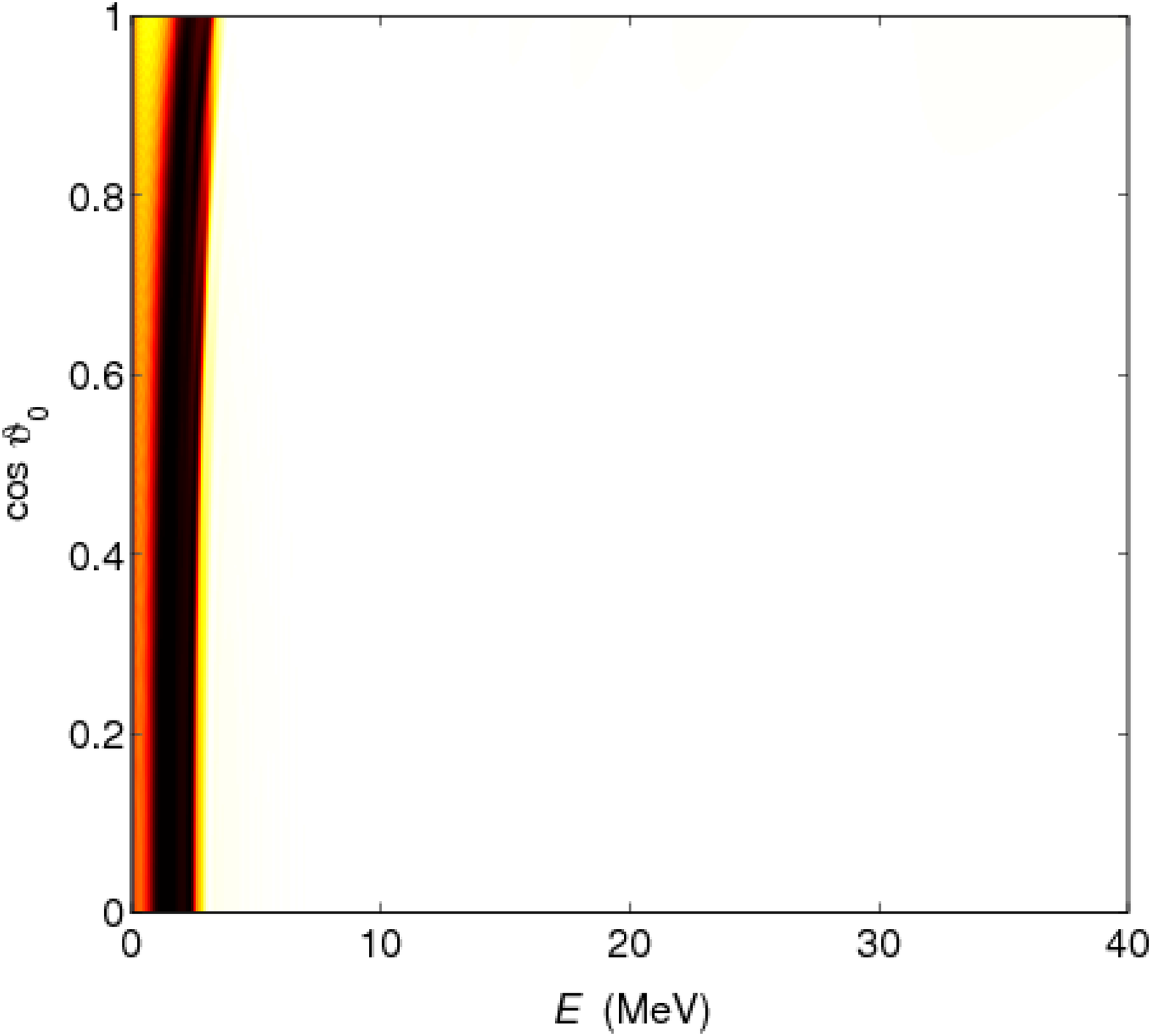} &   
\includegraphics*[scale=0.17, keepaspectratio]{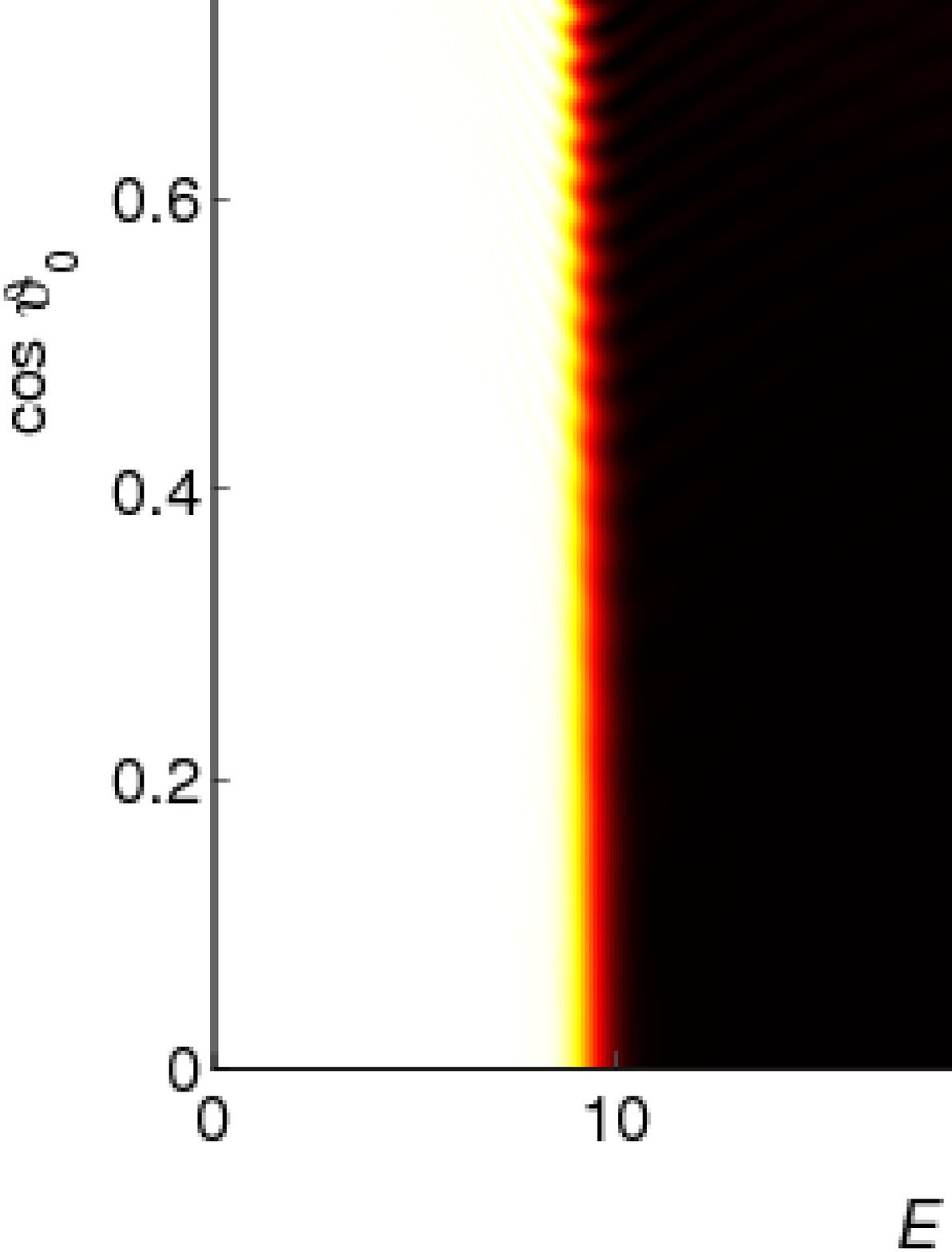}     
\end{array}$
\end{center}
\caption{\label{fig:P-c-E}(Color online)
The neutrino survival probability $P_{\nu\nu}$ as a function of
neutrino emission angle $\vartheta_0$ (relative to the normal at
the emission point on the neutrino sphere) and energy $E$.
The left panel is calculated for a normal
neutrino mass hierarchy with $\thetav=0.01$
and the right panel is for an inverted neutrino mass hierarchy with
$\thetav=10^{-9}$. These results are taken at radius $r=250$ km.
Except for $\thetav$, all parameters are the same as those 
for Fig.~3 of Ref.~\cite{Duan:2006jv}.}
\end{figure*}

In the inverted neutrino mass hierarchy case,
 flavor transformation is suppressed
when $n_e$ and neutrino fluxes are high.
As $\nu_e$ are dominant in supernovae, the bipolar system of
neutrinos and antineutrinos resembles a gyroscopic pendulum near  
its highest point (displacement angle equal to $\pi$) in flavor space.
For a simple bipolar system initially consisting of mono-energetic
$\nu_e$ and $\bar\nu_e$, the analogy is exact, with the initial displacement
angle being $\pi-2\thetav$. When the total neutrino flux
decreases below some critical value, the flavor
pendulum evolves away from its maximum displacement
\cite{Hannestad:2006nj,Duan:2007mv}.
Its nutation then pushes neutrinos and antineutrinos into 
the collective precession mode. 
The presence of a matter field does not change this nutation 
qualitatively, but effectively reduces the
mixing angle \cite{Duan:2005cp,Hannestad:2006nj}.
Therefore, $\Lm$ is essentially unchanged before the precession 
mode begins. As a result, when $n_e$ and neutrino fluxes become 
small, $f_{\nu_1(\bar\nu_1)}(E)$ and $f_{\nu_3(\bar\nu_3)}(E)$ 
are related simply  through Eq.~\eqref{eq:step}
to the initial neutrino energy spectra at the neutrino 
sphere. Specifically, for $\thetav\ll1$
we have 
\begin{equation}
\begin{split}
\Lm&\simeq\int_0^{\EC} [f_{\nu_e}(E)-f_{\nu_{\tau^*}}(E)]\,\ud E
\\
&+\int_{\EC}^\infty [f_{\nu_{\tau^*}}(E)-f_{\nu_e}(E)]\,\ud E
+\frac{n_{\bar\nu_e}-n_{\bar\nu_{\tau^*}}}{n_\nu^\mathrm{tot}}
\\
&\simeq\frac{n_{\nu_e}-n_{\nu_{\tau^*}}-n_{\bar\nu_e}+n_{\bar\nu_{\tau^*}}}
{n_\nu^\mathrm{tot}},
\end{split}
\label{eq:Lf}
\end{equation}
where, e.g., $f_{\nu_e}(E)$ and $n_{\nu_e}(E)$ are
the initial spectrum (normalized to $n_\nu^\mathrm{tot}$)
and number density, respectively, of $\nu_e$ at the neutrino sphere.
 The transition energy $\EC$ can then be found
from Eq.~\eqref{eq:Lf}.

The conservation of $\Lm$ can also be used to find $\EC$ for
the normal mass hierarchy case. However, in this case, $\Lm$
cannot be related simply to the initial neutrino spectra at the
neutrino sphere through Eq.~\eqref{eq:step}. This is because
there is a resonance in the quasi-static MSW-like solution initially
followed by neutrinos and antineutrinos. For example,
in the large neutrino luminosity limit, all neutrinos and antineutrinos 
are synchronized \cite{Pastor:2001iu} and
 experience simultaneously an MSW-like 
resonance near the radius where a single $\nu_e$ with a representative energy 
$\Esync$ would encounter a conventional MSW resonance \cite{Pastor:2002we}.
During this MSW-like evolution, $\Lm$ is decreased. 
Using the initial spectra for supernova neutrinos and 
Eqs.~\eqref{eq:step} and \eqref{eq:Lm},
it can be shown that the less $\Lm$ is reduced, 
the smaller $\EC$ becomes. 
If $\thetav$ is tiny and/or the neutrino luminosities
are not large enough, the MSW-like conversion will be non-adiabatic, 
and $\Lm$ will change very little. In this case,
$\EC\rightarrow 0$ and the stepwise nature of the swapping of
$\nu_e$ and $\nu_{\mu,\tau}$ spectra becomes unobservable.
On the other hand, given sufficiently large values of
$\thetav$ and/or neutrino luminosities, MSW-like flavor
conversion can be adiabatic and efficient for a large
range of neutrino energies. This can engender nearly complete
swapping of the entire neutrino and antineutrino spectra and,
consequently, $\Lm$ can retain its magnitude but reverse its sign.
In this case, $\EC$ for the subsequent stepwise spectral swapping
 is roughly the same as in the inverted
neutrino mass hierarchy case.

Though the transition energy $\EC$
is sensitive to $\thetav$ in the normal mass hierarchy case,
it appears to be essentially independent of $\thetav$ in the 
inverted mass hierarchy case. We have carried out multi-angle 
simulations under the same conditions
as those discussed in Refs.~\cite{Duan:2006jv,Duan:2006an} except
with smaller $\thetav$. The probability 
$P_{\nu_e\nu_e}(E,\vartheta_0)$  at radius $r=250$~km
is plotted in Fig.~\ref{fig:P-c-E}
for both a normal mass hierarchy case ($\thetav=0.01$,
left panel) and an inverted mass hierarchy case ($\thetav=10^{-9}$,
right panel). Here $\vartheta_0$ is the angle between the propagation 
direction of the neutrino and the normal at its emission position
on the neutrino sphere (see Fig.~1 of Ref.~\cite{Duan:2006jv}).
A comparison of the results shown in Fig.~\ref{fig:P-c-E}
with those for $\thetav=0.1$ shown in Fig.~3 of Ref.~\cite{Duan:2006jv} 
is revealing. In the normal neutrino mass hierarchy case,
$\EC$ decreases from $\sim10$ MeV to $\sim3$ MeV
as $\thetav$ is reduced from $0.1$ to $0.01$. However,
for the inverted neutrino mass hierarchy case, 
$\EC$ is essentially unchanged as $\thetav$ is decreased by
8 orders of magnitude. 
The value $\EC\simeq 8.4$ MeV 
calculated from Eq.~\eqref{eq:Lf} for this case agrees very 
well with the numerical results. 
We note that $\EC$ has a slight dependence on
neutrino trajectory ($\cos\vartheta_0$) in the multi-angle simulations
for $\thetav\ll 0.1$.

\begin{figure}
\begin{center}
\includegraphics*[width=\myfigwid, keepaspectratio]{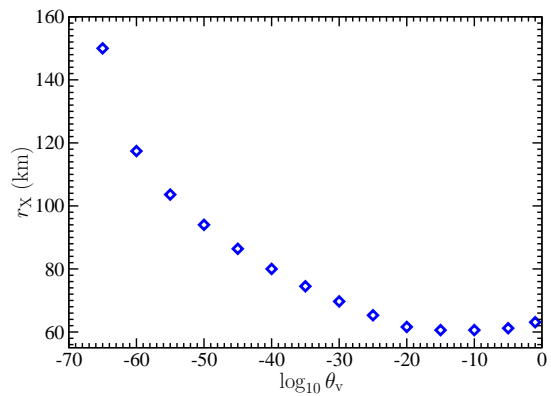}
\end{center}
\caption{\label{fig:rX}%
Single-angle simulation results for the dependence 
of $r_\mathrm{X}$ on $\thetav$ in the inverted neutrino
mass hierarchy case. Here $\thetav$ is the effective
$2\times2$ vacuum mixing angle ($\simeq\theta_{13}$),
and $r_\mathrm{X}$ is the radius where the energy-averaged
value of $P_{\nu_e\nu_e}$ drops below $0.9$. Except for $\thetav$,
all parameters are the same as those for Fig.~2 of Ref.~\cite{Duan:2006jv}.}
\end{figure}

The apparent insensitivity of $\EC$ to $\thetav$ in the inverted
neutrino mass hierarchy case requires discussion.
We note that the neutrino system transitions from the MSW-like
evolution to the collective precession mode through nutation. If the
system does not develop significant nutation while it is
in the collective flavor transformation regime, it will not enter
the collective precession mode, and therefore, stepwise spectral 
swapping will not occur. For a uniform and isotropic gas of 
mono-energetic neutrinos initially in pure $\nu_e$ and
$\bar\nu_e$ states, the nutation timescale is
$T_\mathrm{nut}\sim-\ln\thetav$ in the inverted mass hierarchy case
 \cite{Hannestad:2006nj}. Estimating the nutation timescale
for realistic supernova neutrino systems is problematic, partly because 
this quantity depends on time-varying electron and neutrino number
densities \cite{Kostelecky:1994dt,Duan:2005cp,Hannestad:2006nj}.
In Fig.~\ref{fig:rX} we plot as a function of $\thetav$
the radius $r_\mathrm{X}$ (as defined in Ref.~\cite{Duan:2006an}) 
where the energy-averaged value of
$P_{\nu_e\nu_e}$ drops below $0.9$  and 
significant nutation develops in our single-angle simulations for
the inverted mass hierarchy. Our single-angle calculations suggest
that the onset of significant nutation is nearly independent of
$\thetav$ as it is decreased from 0.1 to
$\simeq 10^{-20}$. As $\thetav$ is decreased further,
$r_\mathrm{X}$ begins to increase. We expect that for sufficiently
small $\thetav$, the onset of significant nutation is pushed to so
large a radius that the corresponding neutrino number density 
becomes too low to generate any collective flavor evolution or
stepwise spectral swapping.

% Neutrino signals from a Galactic supernova are
% led by a short intense burst of  neutrinos which are 
% predominantly $\nu_e$'s emitted when the shock breaks through the neutrino
% sphere. 
Just after the bounce of the supernova core,
when the supernova shock breaks through the neutrino sphere,
there is a brief intense burst of neutrinos
which are emitted predominantly as $\nu_e$'s.
Lacking in $\bar\nu_e$'s \cite{Kachelriess:2004ds}, this burst
is not likely to be affected by stepwise spectral swapping
at the $\delta m_\mathrm{atm}^2$ scale
(but see Ref.~\cite{Duan:2007sh}). Later, both neutrinos and
antineutrinos are emitted and they form a bipolar system.
If $\theta_{13}$ is not too small,
for the normal mass hierarchy case, stepwise spectral swapping
at the $\delta m_\mathrm{atm}^2$ scale
can be observable at
late times when matter 
has been sufficiently condensed toward the proto-neutron star.
For the inverted mass hierarchy case, however, stepwise spectral swapping
occurs even at early times \cite{Duan:2005cp,Fogli:2007bk}
for essentially any nonvanishing $\theta_{13}$. In light of 
the insensitivity of $\EC$ to $\theta_{13}$ in the latter case,
supernova neutrino signals can offer a unique probe of
the neutrino mass hierarchy even for $\theta_{13}$ too small
to be measured by conventional neutrino oscillation experiments.

\begin{acknowledgments}
This work was supported in part by 
NSF grant PHY-04-00359,
the TSI collaboration's DOE SciDAC grant at UCSD, 
DOE grant DE-FG02-87ER40328 at UMN,
and an IGPP/LANL mini-grant.
This work was also supported in part by the LDRD Program
and Open Supercomputing at LANL, and by
NERSC through
the TSI collaboration using Bassi. We thank A.~Burrows,
A.~Mezzacappa and A.~Mirizzi for valuable conversations.
\end{acknowledgments}

\bibliography{ref}

\end{document}